\RequirePackage{fix-cm}
\documentclass[smallextended]{svjour3}       
\smartqed  
\usepackage{graphicx}
\usepackage{multirow}
\usepackage{epsfig}
\journalname{}

\title{Proofs of the Kochen-Specker theorem based on a system of three qubits}

\author{Mordecai Waegell and P.K. Aravind }

\authorrunning{M.Waegell, P.K. Aravind}

\institute{M.Waegell, P.K. Aravind \at
Physics Department, Worcester Polytechnic Institute, Worcester, MA 01609, U.S.A.\\
\email{caiw@wpi.edu, paravind@wpi.edu}}

\date{\today}

\begin{document}
\maketitle
\begin{abstract}
A number of new proofs of the Kochen-Specker theorem are given based on the observables of the three-qubit Pauli group. Each proof is presented in the form of a diagram from which it is obvious by inspection. Each of our observable-based proofs leads to a system of projectors and bases that generally yields a large number of ``parity proofs" of the Kochen-Specker theorem. Some examples of such proofs are given and some of their applications are discussed.
\end{abstract}

\section{\label{sec:Intro}Introduction}
In the 1990s Greenberger, Horne and Zeilinger (GHZ) \cite{GHZ}, Peres \cite{Peres1991} and Mermin \cite{Mermin1993} gave a number of proofs of the Kochen-Specker (KS) \cite{KS1967} and$/$or Bell \cite{Bell1964} theorems using systems of two to four qubits. The work of GHZ was remarkable for showing that Bell's theorem could be proved ``without probabilities", while the works of Peres \cite{Peres1991}, Mermin \cite{Mermin1993} and Kernaghan and Peres \cite{KP1995} were notable for giving simple proofs of the KS theorem based on systems of two or three qubits. The proofs of the KS theorem in \cite{Peres1991,Mermin1993} were later extended into an inequality-free proof of Bell's theorem by using a suitable entangled state shared between two distant observers \cite{Cabello2001}. Generalizations of Bell proofs to higher dimensional state spaces and more than two observers have been made\cite{Bell2006}. In parallel with these developments, there has been a renewed interest in the KS theorem and its role in establishing contextuality. A new route to this goal was pointed out by Cabello \cite{Cabello2008}, who showed how any proof of the KS theorem can be converted into an inequality that is satisfied by a noncontextual theory but violated by quantum mechanics. Experiments carried out on a variety of four-state systems \cite{Kirchmair,Bartosik,Amselem,Moussa} have shown violations of such Cabello-like inequalities and thus established quantum contextuality. In another notable development, Klyachko et al \cite{Klyachko} derived a pentagon inequality for three-state systems that provides perhaps the simplest demonstration of contextuality in the lowest dimension (three) in which it is possible. This approach was later extended to four and higher state systems \cite{Liang,Bengtsson}. Recently Yu and Oh \cite{YuOh} have shown, in a different way, that KS sets are not needed to establish contextuality. On the more formal side, an interesting connection between KS proofs and ``logical Bell inequalities" has been made in \cite{Abram}.\\

The purpose of this paper is to show that the proofs of the KS theorem given in \cite{Mermin1993,KP1995} are just the tip of an iceberg and that there are a large number of other proofs of this kind in the three-qubit system. This set of proofs is interesting for the following reasons:\\

(1) Each proof can be presented in the form of a diagram from which it is obvious by inspection.\\

(2) Each of our observables-based proofs can be used to generate a large number of parity proofs of the KS theorem. Only one parity proof was known previously in a three-qubit system, namely, the Kernaghan-Peres proof \cite{KP1995} involving 36 rank-1 projectors and 11 bases. We have found a large number of new proofs, including several that are more economical than the Kernaghan-Peres proof in that they involve both a smaller number of projectors and bases.\\

(3) Our proofs can be translated into inequalities of the type proposed by Cabello \cite{Cabello2008} for establishing quantum contextuality.\\

(4) The techniques of this paper can be generalized to obtain KS proofs for $N$-qubit systems, for $N \geq 4$ (a matter we will elaborate on elsewhere). This is of interest because it permits the construction of compact sets of KS vectors in high dimensions, and would therefore be a useful adjunct to other known methods of constructing such sets \cite{CEG2005,ZP}.\\

The plan of this paper is as follows: Sec. \ref{sec:2} reviews the proofs of the KS theorem based on two qubits and shows how they can be represented in the form of diagrams; Sec. \ref{sec:3} presents our new KS proofs based on three qubits in the form of diagrams; Sec \ref{sec:4} demonstrates how our observable-based proofs in Sec. \ref{sec:3} can be made to yield parity proofs of the KS theorem; and finally Sec \ref{sec:5} contains some concluding remarks.

\section{\label{sec:2}KS proofs based on the two-qubit Pauli group}

The two-qubit Pauli group has 15 nontrivial observables which we will denote by the symbols $XY$, $IZ$, etc., where the first and second letters in each symbol refer to the Pauli operators and the identity ($X,Y,Z$ and $I$) of the two qubits. Peres \cite{Peres1991} and Mermin \cite{Mermin1993} arranged nine of these observables in the form of a 3 x 3 square that they used to give (somewhat different) proofs of the KS theorem. Figure 1 shows the Peres-Mermin square in two alternative guises: as a square and a wheel. These diagrams (as well as all the later ones of this paper) have been designed to make the KS proofs in them come across as simply as possible. The rules underlying the construction of our diagrams are as follows:\\

R1. Each observable is represented within a circle.\\

R2. Members of a set of mutually commuting observables are connected by a line (that can be straight or curved).\\

R3. The product of the observables in any mutually commuting set in our diagrams is always {+\bf I} or {$-$\bf I}, where {\bf I} is the identity operator in the space of all the qubits. We will term such a commuting set an ID (a contraction for ``Identity") and also characterize it as positive or negative according as the product of the observables in it is {+\bf I} or {$-$\bf I}. Positive IDs are represented by thin lines and negative IDs by thick ones.\\

The Peres-Mermin square, in both the forms shown in Fig. 1, illustrates these rules, as the reader can readily check. The single negative ID in each case (a line in the square and a circle in the wheel) is indicated by a thick line. We will term the IDs that occur in the Peres-Mermin square ID3s because they each involve three mutually commuting observables whose product is the identity (up to a sign). The only IDs that can ever occur in any three-qubit KS proof (and therefore the only IDs that will show up in our diagrams) are ID3s and ID4s (the latter involves a set of four mutually commuting observables whose product is {+\bf I} or {$-$\bf I}).\\

\begin{figure}
\centering
\begin{tabular}{cc}
\epsfig{file=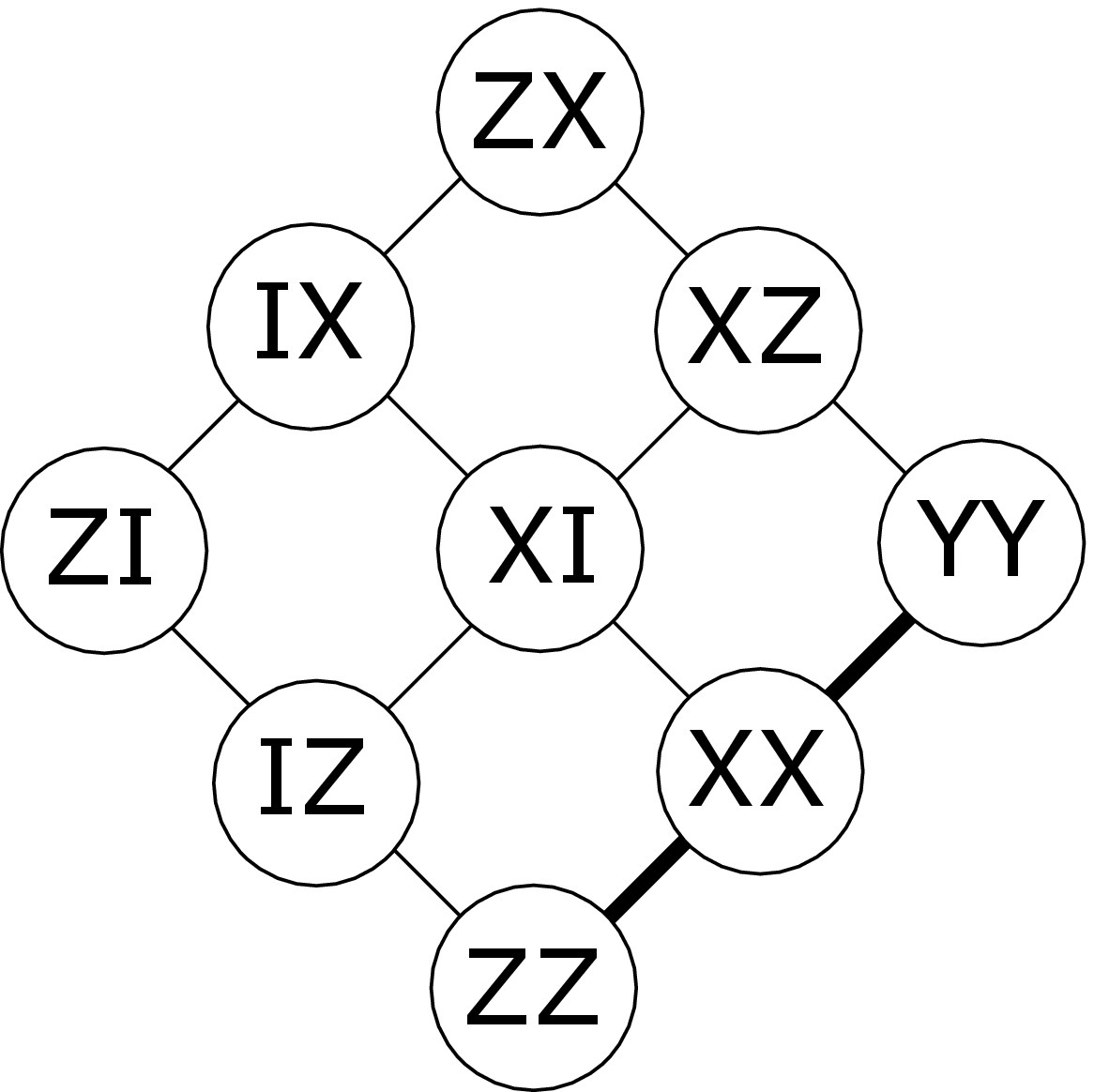,width=0.4\linewidth,clip=} &
\epsfig{file=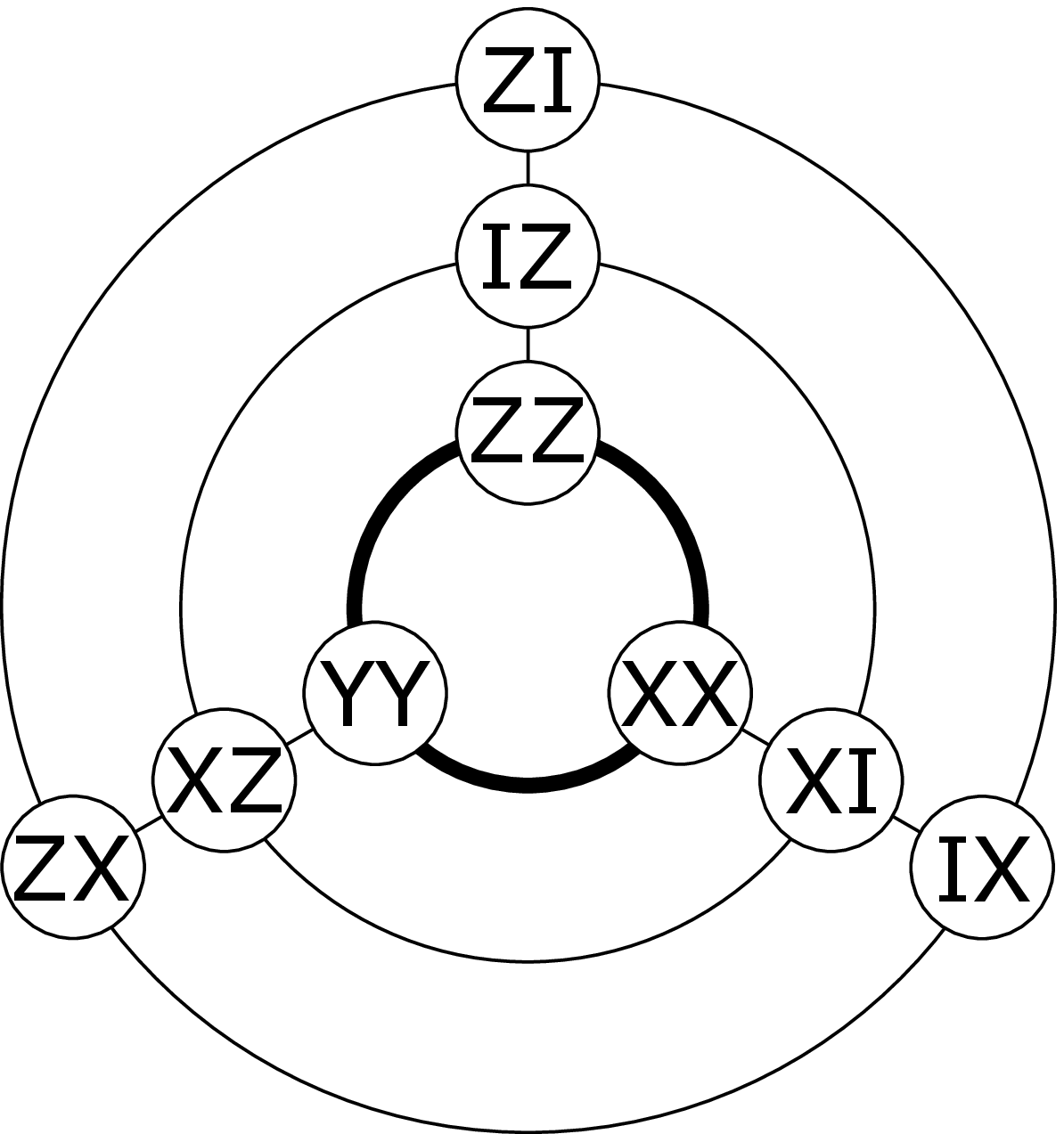,width=0.4\linewidth,clip=} \\
\end{tabular}
\caption{The two-qubit Peres-Mermin square, symbol $9_{2}$-$6_{3}$, in the form of a square (left) and a wheel (right).}
\label{Fig.1}
\end{figure}

In order to check that any of our diagrams provides a proof of the KS theorem, it is only necessary to verify that (a) each observable occurs in an even number of IDs, and (b) there are an odd number of negative IDs (or thick lines) in the diagram. To see why these two properties guarantee a KS proof, note that the square of any of our observables is always the identity, implying that their eigenvalues are $\pm 1$. Now, a noncontextual hidden variables theory is required to assign the value $+1$ or $-1$ to each of the observables in such a way that both the following conditions are satisfied: (A) Product Rule -- the product of the values assigned to the observables in a positive or negative ID must be equal $+1$ or $-1$, respectively; and (B) Noncontextuality -- the value assigned to an observable must be independent of context, i.e., it must be the same no matter which ID the observable is considered a part of. The impossibility of a value assignment satisfying both conditions (A) and (B) is proved by any diagram possessing both properties (a) and (b). To see this, let $v_{\alpha}$ be the product of the values of the observables in the ID indexed by $\alpha$, and consider the product of the products, $P = \prod v_{\alpha}$, taken over all the IDs. On the one hand, conditions A and B require $P = -1$ because the diagram obeys property (b); but, on the other hand, they also require $P = +1$ because the diagram obeys property (a). This contradiction shows that noncontextual value assignments are impossible and hence proves the KS theorem.\\

It is useful to introduce symbols for our diagrams, as a way of summarizing some of their key features. We will use the symbol $9_{2}$-$6_{3}$ for the Peres-Mermin square because it consists of 9 observables and 6 IDs, with each observable occurring in two IDs and each ID consisting of three observables. These numbers are not independent but obey the relation $9 \times 2 = 6 \times 3$, which expresses the fact that the total count of observables over all the IDs is a multiple (here two) of their actual number. Some of our later diagrams will have more complicated symbols of the form $A_{i}B_{j}$-$N_{3}M_{4}$, which expresses the fact that there are $A$ observables of multiplicity $i$, $B$ observables of multiplicity $j$, $N$ ID3s and $M$ ID4s in the diagram. Again these numbers are not independent but obey the constraint $A\times i + B \times j = 3N +4M$. The fact that a diagram is consistent with property (a) required for a KS proof can be checked by confirming that all the subscripts in the first half of its symbol are even. Property (b) is not reflected in the symbol but can be checked only by looking at the diagram. While a diagram has a unique symbol associated with it, the reverse is not true (i.e., that there is just one proof, up to unitarity, associated with any symbol). Figure 7 of the next section shows two inequivalent three-qubit diagrams having the same symbol.\\

\begin{figure}[htp]
\begin{center}
\includegraphics[width=.90\textwidth]{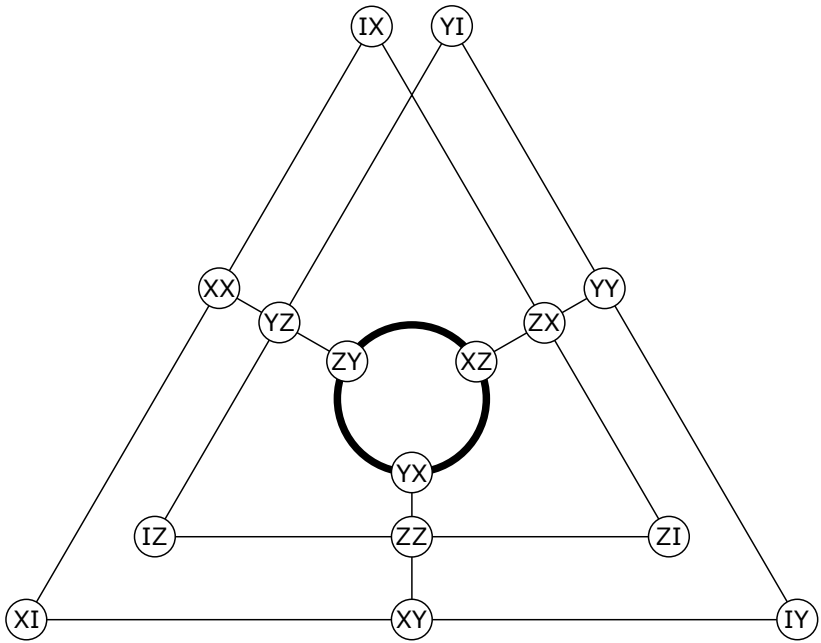}
\end{center}
\caption{The two-qubit $15_{2}$-$10_{3}$ proof.}
\label{Fig.2}
\end{figure}

Aside from the Peres-Mermin square, the only other KS proof that can be built out of the observables of the two-qubit Pauli group is the one shown in Fig. 2. It was presented for the first time in \cite{Waegell2011}, but without the diagram. The reader can readily check that this diagram satisfies properties (a) and (b). The symbol for this diagram (or the associated proof) is $15_{2}$-$10_{3}$.\\

Peres \cite{Peres1991} used the diagram of Fig. 1 to construct 24 real four-dimensional vectors, that form 24 bases, and used them to give a proof of the KS theorem. It was shown later \cite{Kernaghan1994,Cabello1996,Aravind2000,Pavicic2010,Waegell2011b} that this set contains many smaller sets of rays and bases that provide parity proofs of the KS theorem. A convenient way of characterizing a parity proof is by a symbol like R-B, where the integers R and B denote the number of rays and bases in the proof. The 24-24 Peres system has proofs of the four types 18-9, 20-11, 22-13 and 24-15 in it. However each of these types has many replicas under symmetry and the total number of all the distinct proofs is $2^{9}=512$. The diagram of Fig. 2 gives rise to a system of 40 rays and 40 bases that contain six different types of parity proofs (30-15,32-17,34-19,36-21,38-23 and 40-25), with the total number of proofs (when all replicas under symmetry are taken into account) being $2^{15}=32768$ \cite{Waegell2011}.\\

This completes our survey of KS proofs in the two-qubit system. We now pass to the three-qubit system, where we have several new results to report.

\section{\label{sec:3}KS proofs based on the three-qubit Pauli group}

Figure 3 shows two slightly different versions of a KS proof based on three qubits. The one on the left, due to Mermin \cite{Mermin1993}, has just one negative ID4 while the one on the right has five. The latter proof is clearly more difficult to realize experimentally, but we nevertheless mention it as a possibility allowed by the Pauli group.\\

\begin{figure}
\centering
\begin{tabular}{cc}
\epsfig{file=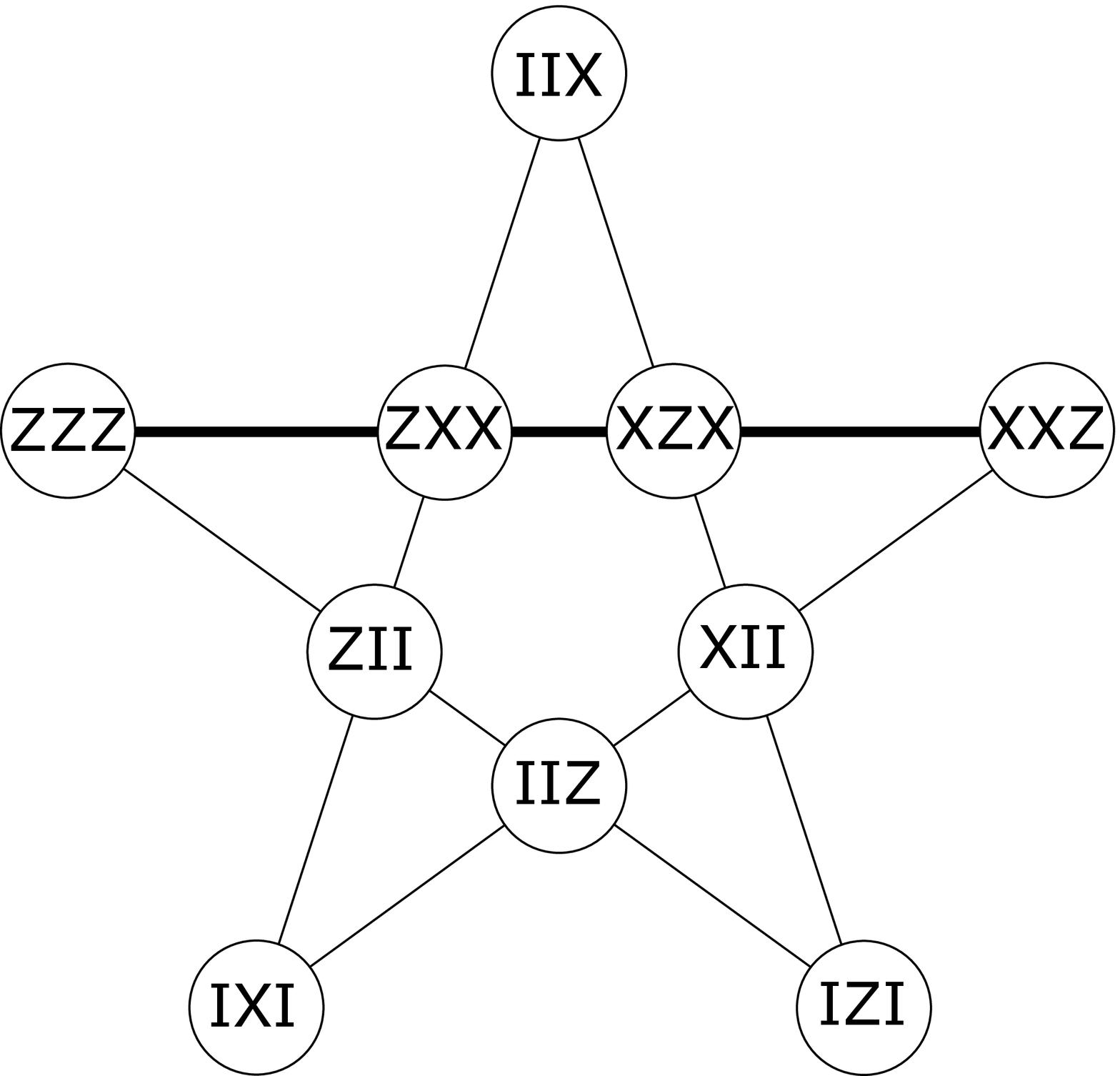,width=0.5\linewidth,clip=} &
\epsfig{file=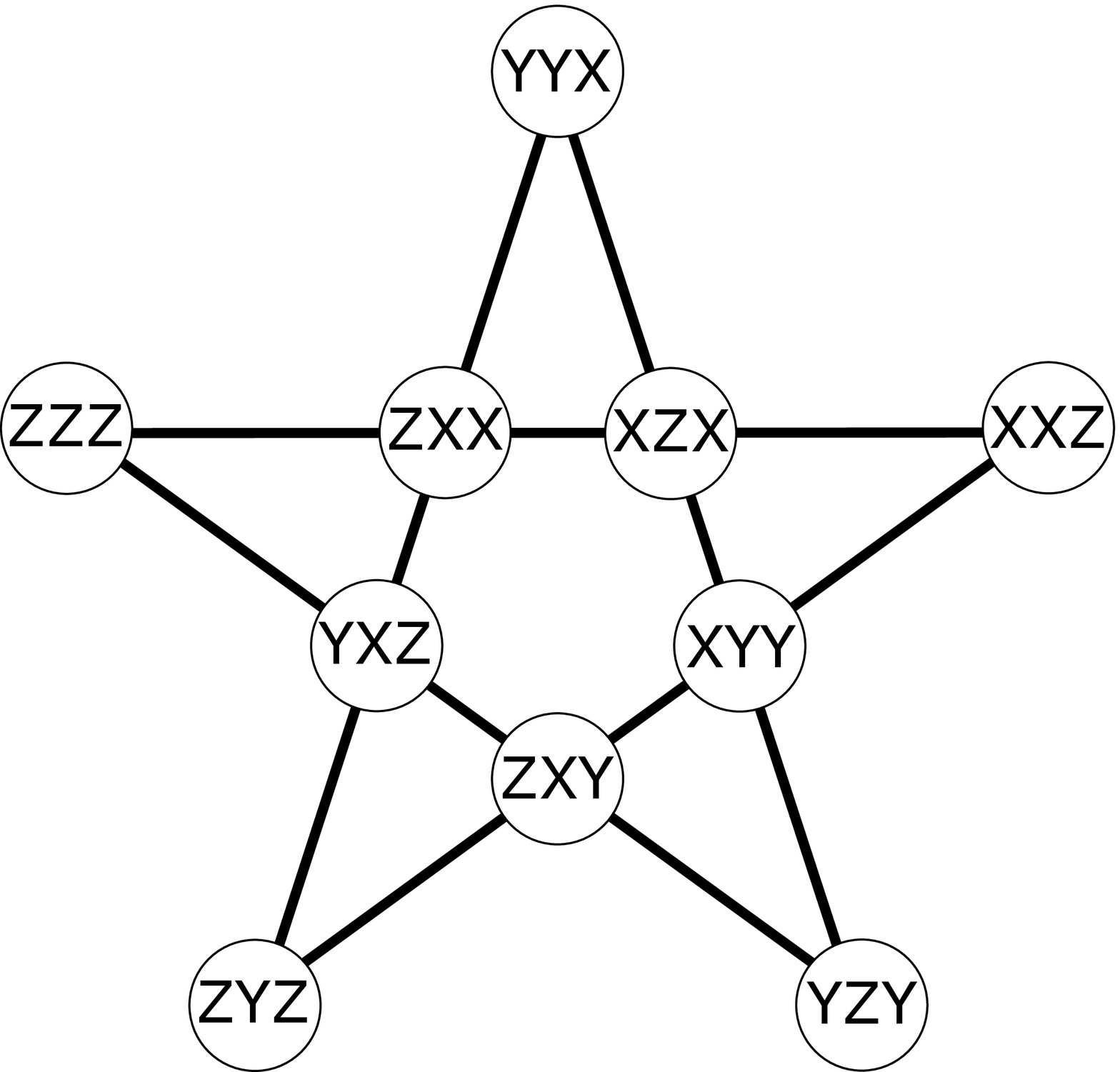,width=0.5\linewidth,clip=} \\
\end{tabular}
\caption{The three-qubit GHZ-Mermin pentagram (left) and a variation of it involving only negative ID4s (right). Both diagrams have the symbol $10_{2}$-$5_{4}$.}
\label{Fig.3}
\end{figure}

Figure 4 shows a three-qubit generalization of the Peres-Mermin square, both in the form of a Square and a Wheel. A hierarchy of new diagrams (or proofs) can be created by replacing one, two, or all three circles in the Wheel by triangular arrays of observables with ID3s along their edges. The diagram that results when all three circles in the Wheel are replaced in this manner is shown in Fig. 5.\\

\begin{figure}
\centering
\begin{tabular}{cc}
\epsfig{file=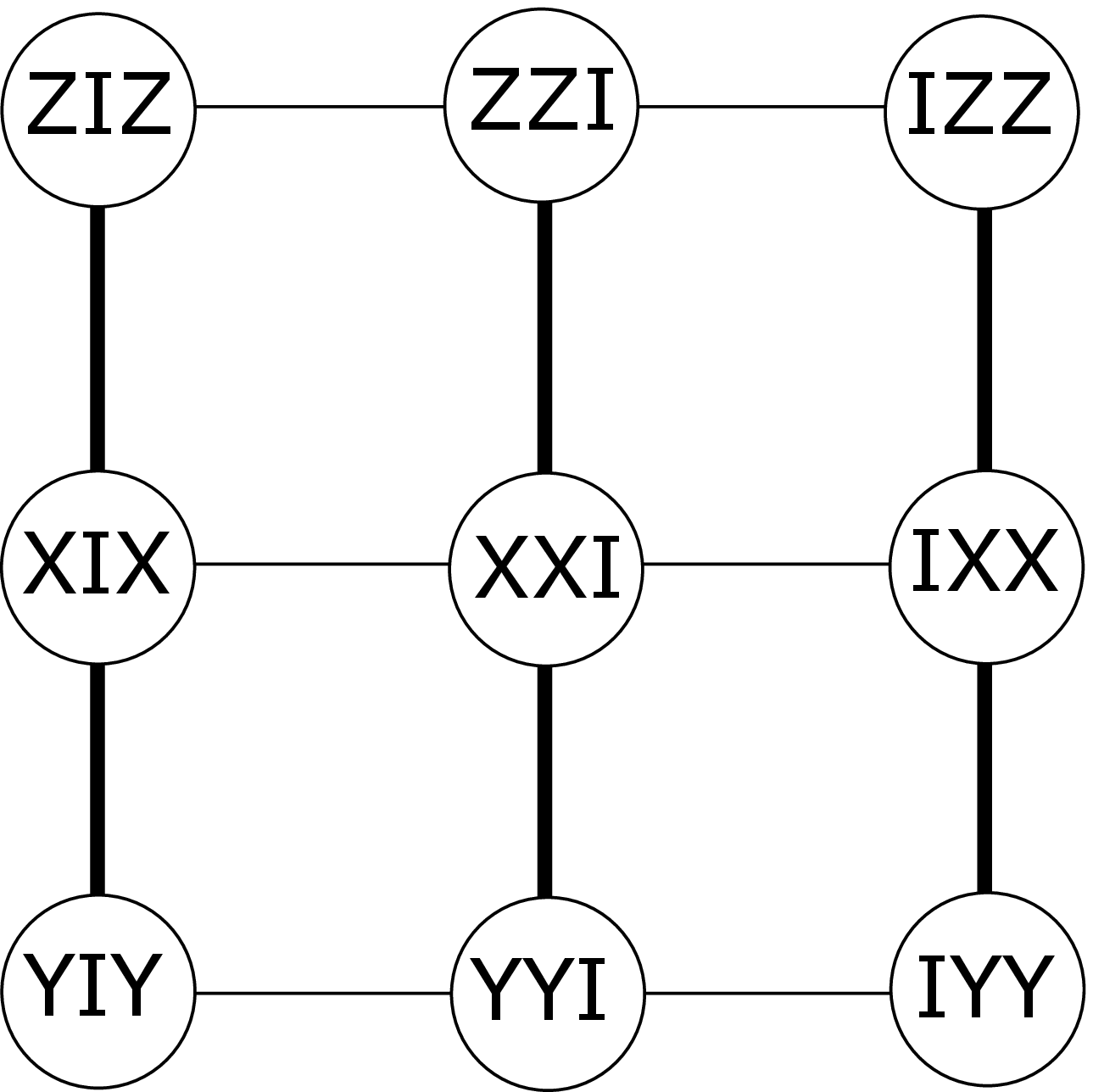,width=0.4\linewidth,clip=} &
\epsfig{file=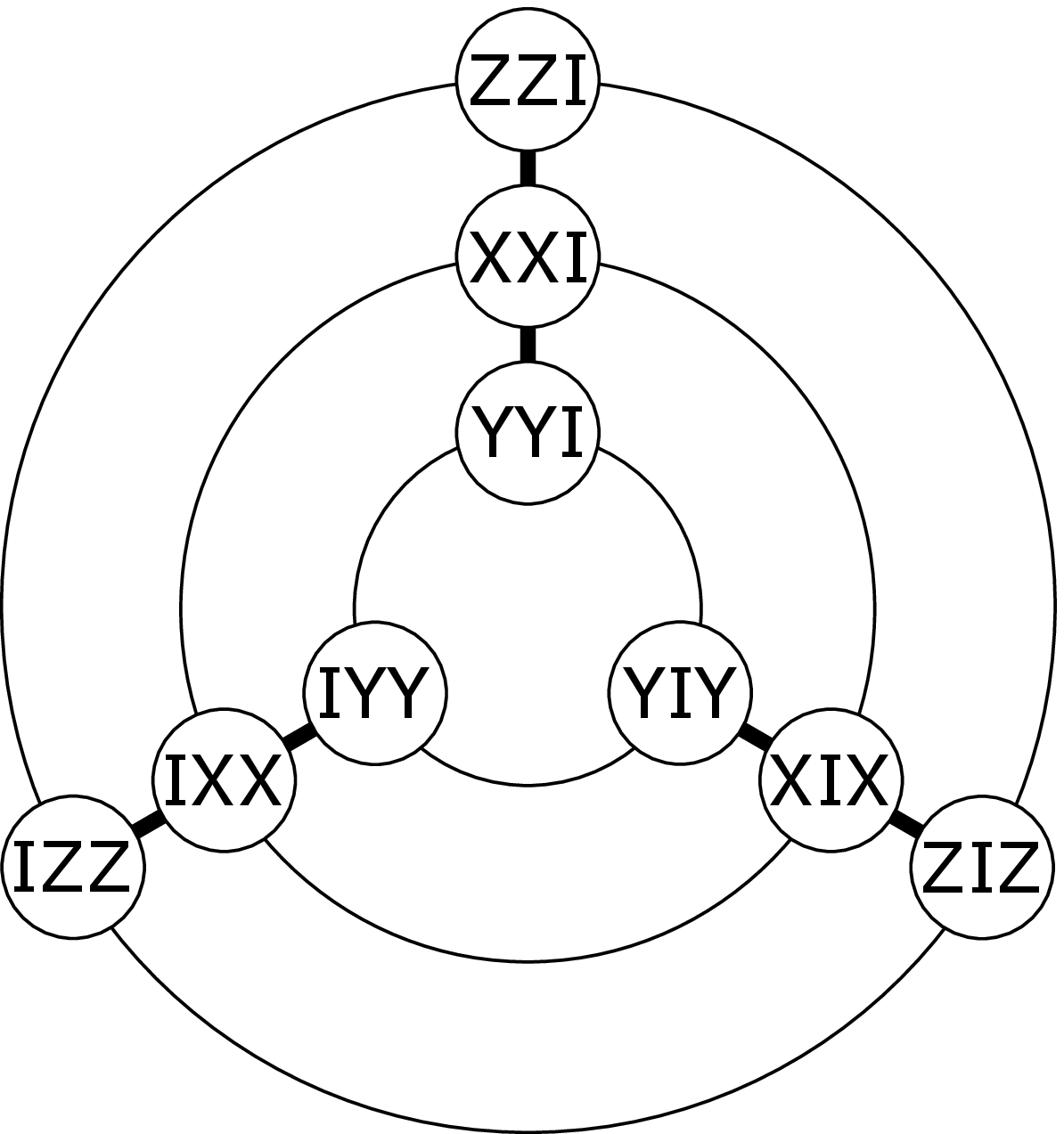,width=0.4\linewidth,clip=} \\
\end{tabular}
\caption{The three-qubit Peres-Mermin square, symbol $9_{2}$-$6_{3}$, as a Square (left) and a Wheel (right).}
\label{Fig.4}
\end{figure}

\begin{figure}[htp]
\begin{center}
\includegraphics[width=.80\textwidth]{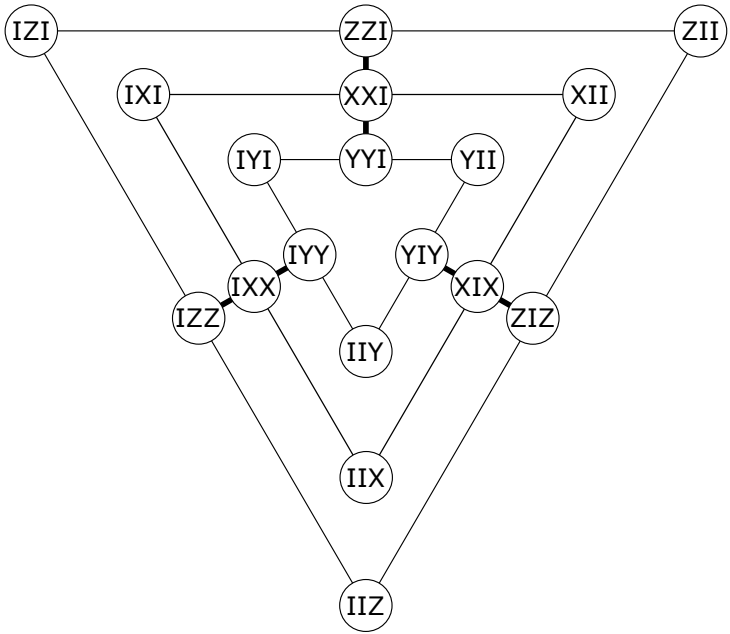}
\end{center}
\caption{The three-qubit diagram, symbol $18_{2}$-$12_{3}$, obtained by replacing all three circles in the Wheel of Fig. 4 by triangles.}
\label{Fig.5}
\end{figure}

Figure 6, which is in the shape of a kite, is the first of our diagrams to have both ID3s and ID4s in it. The two ID4s (one positive and the other negative) share two observables that lie at the tail of the kite. The symbol for this diagram may aid the reader in identifying the IDs in it, and in checking that it does really provide a KS proof. As the diagrams become more complicated, the reader may find that their symbols prove increasingly helpful in sorting out their details.\\

\begin{figure}[htp]
\begin{center}
\includegraphics[width=.40\textwidth]{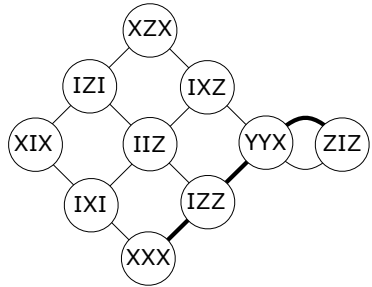}
\end{center}
\caption{The three-qubit diagram Kite diagram, symbol $10_{2}$-$4_{3}2_{4}$.}
\label{Fig.6}
\end{figure}

Figure 7 shows a pair of diagrams obtained by modifying the square framework of Fig. 4 to make room for an additional ID4. Figure 8 shows a couple of other proof-diagrams we have found. We close this tour by stressing that we have not displayed all the diagrams we have found, but only a representative sample. A complete classification of all the inequivalent types would clearly be of interest, but we postpone that analysis to a future date.\\

\begin{figure}
\centering
\begin{tabular}{cc}
\epsfig{file=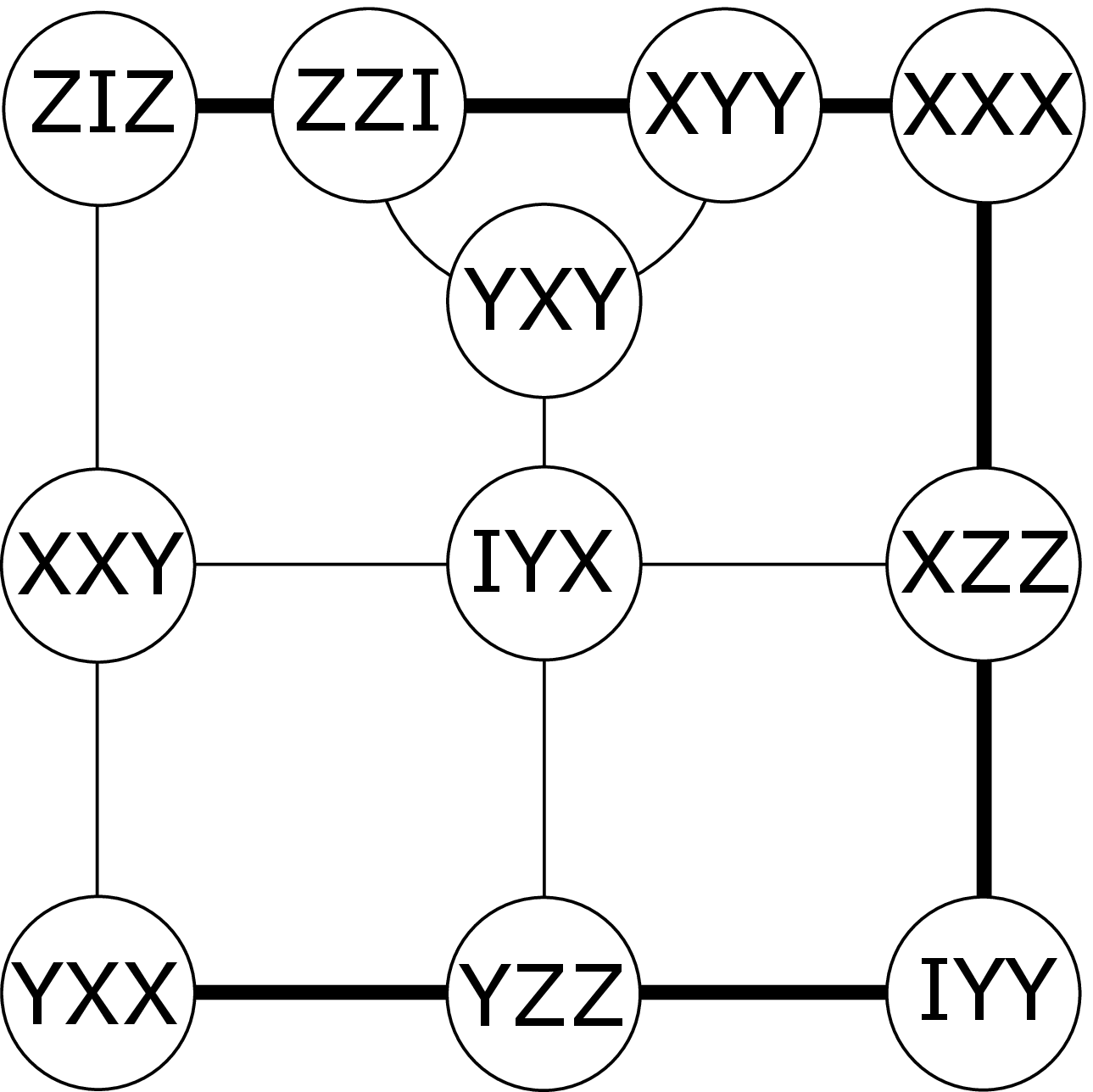,width=0.5\linewidth,clip=} &
\epsfig{file=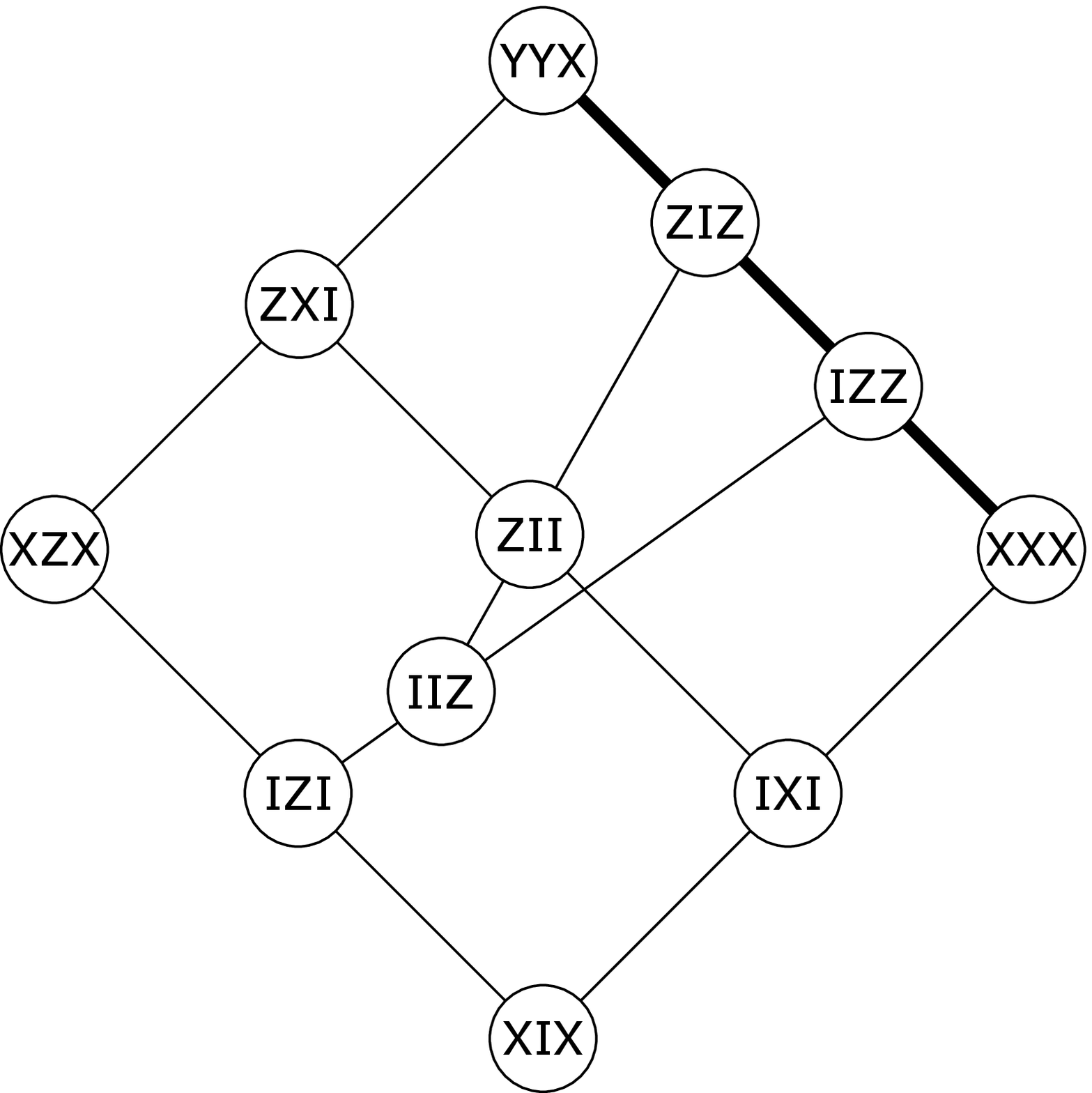,width=0.5\linewidth,clip=} \\
\end{tabular}
\caption{The three-qubit Window (left) and a variation of it (right). Both diagrams have the symbol $11_{2}$-$6_{3}1_{4}$.}
\label{Fig.7}
\end{figure}

\begin{figure}
\centering
\begin{tabular}{cc}
\epsfig{file=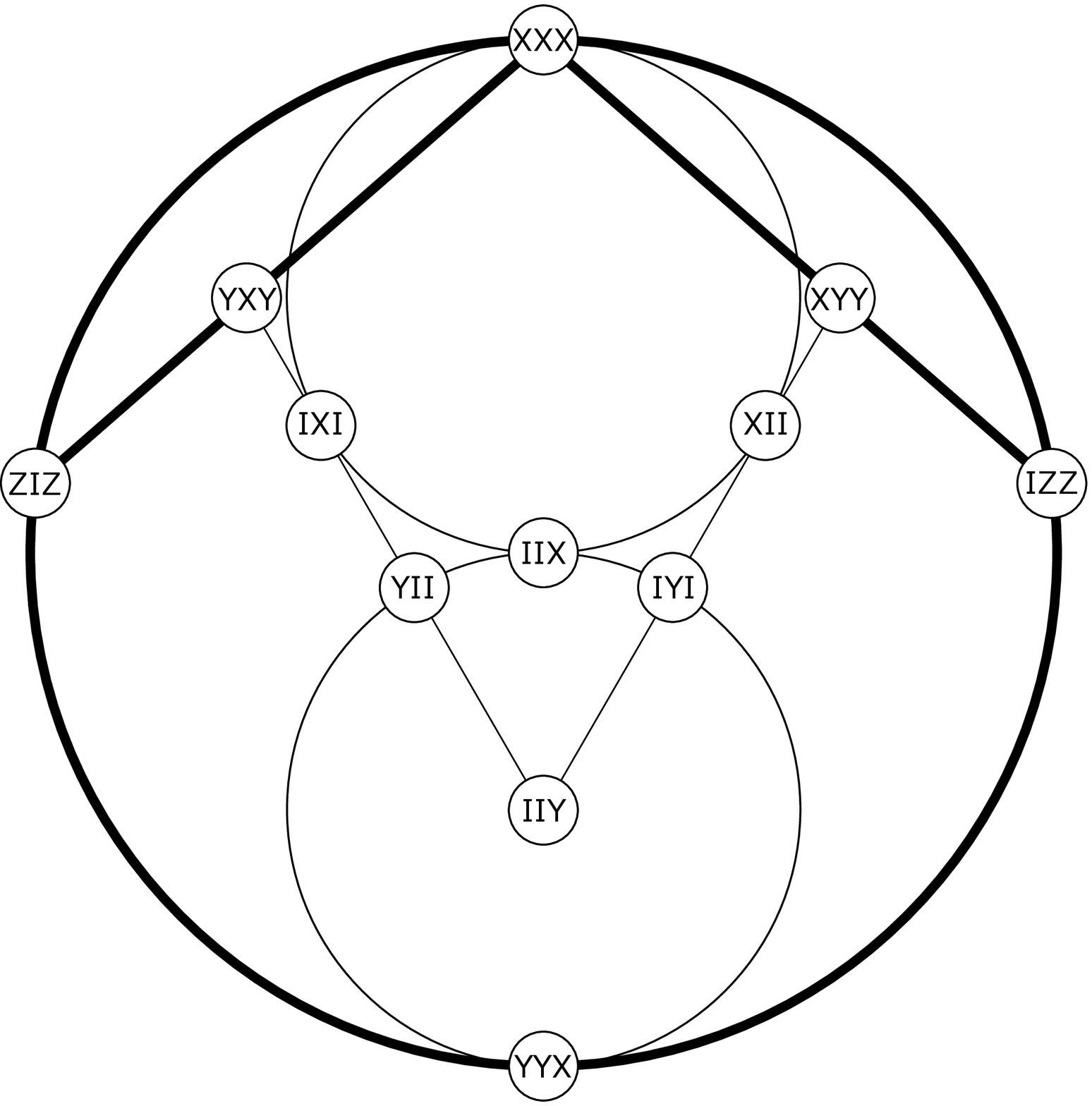,width=0.6\linewidth,clip=} &
\epsfig{file=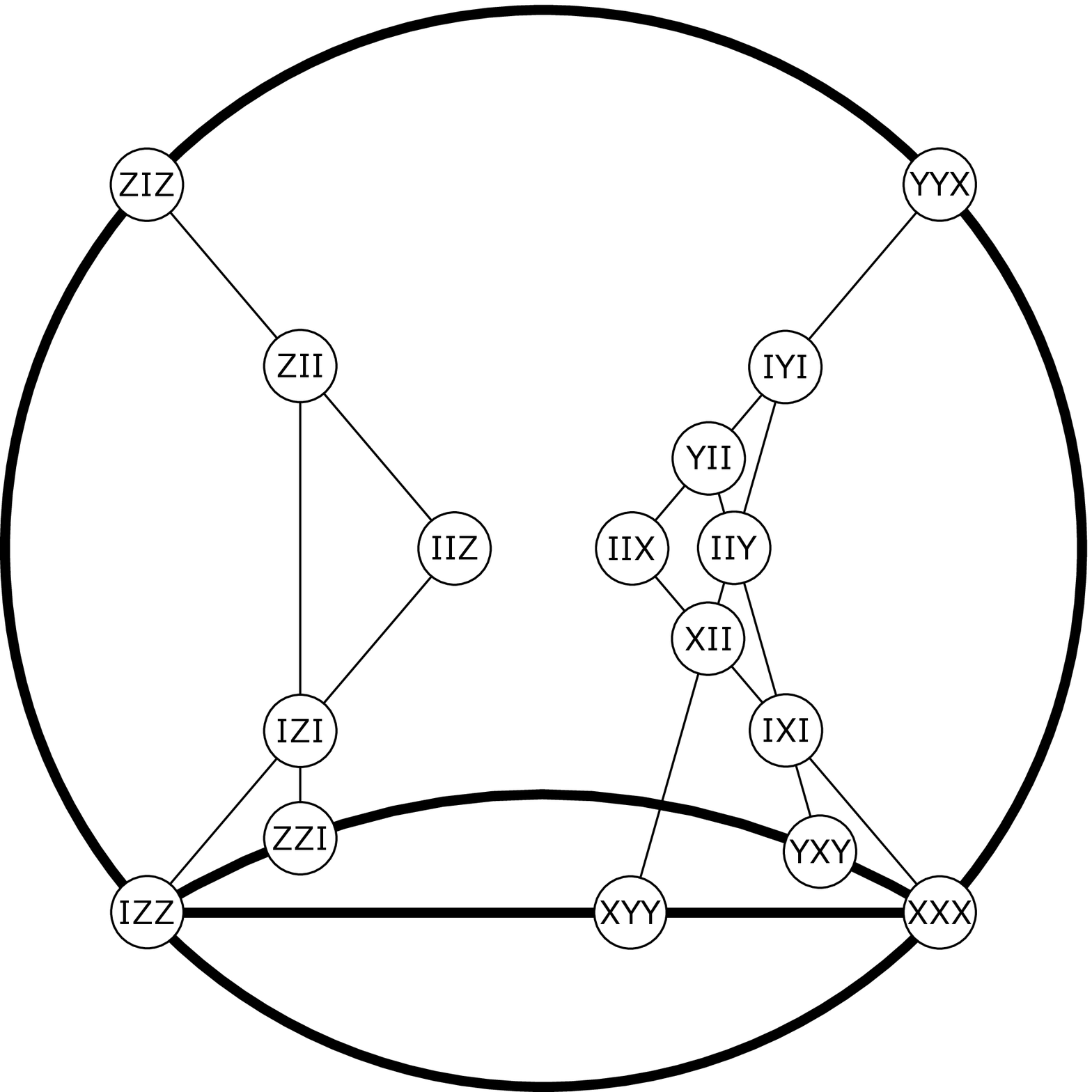,width=0.6\linewidth,clip=} \\
\end{tabular}
\caption{Two three-qubit ``Clock" diagrams, the left one having the symbol $1_{4}11_{2}$-$2_{3}5_{4}$ and the right one the symbol $2_{4}14_{2}$-$4_{3}6_{4}$.}
\label{Fig.8}
\end{figure}

We now say a few words about how we arrived at our diagrams. The three-qubit Pauli group has $4^{3}-1 = 63$ nontrivial observables that form 135 complete sets of commuting observables (of seven members each). These sets have been studied earlier in connection with the construction of maximal sets of mutually unbiased bases \cite{Durt}. An interesting feature of the three-qubit Pauli group, not shared by the two-qubit group, is that a subset of a complete commuting set of observables can have {+\bf I} or {$-$\bf I} as its product. The ID3s and the ID4s are the two subsets of this kind in the three-qubit Pauli group. The strategy we used in constructing our proof-diagrams was to start from a small set of ID3s and/or ID4s and gradually add on others until both properties (a) and (b) were satisfied. It is sometimes possible to create new diagrams by modifying existing ones, as we did in the transition from Fig. 4 to Fig. 5 or between the two diagrams in Fig. 7.\\

We should stress that we have presented only ``critical" diagrams, where by a critical diagram we mean one that cannot be reduced to a simpler one by either discarding some of the IDs and observables or by ignoring one or more qubits in all the observables (thus passing to a smaller qubit system). A couple of examples should help illustrate what we mean. Consider the diagram of Fig. 5, which might not appear to be critical because it would seem to be reducible to the diagram of Fig. 4 if one deletes the observables at the corners of the three triangles. However this objection is not valid because one would have to restore the IDs corresponding to the triads of observables at the corners of the triangles, which show up as circular loops in Fig. 4, in order to recover that diagram. As a second example, consider the three-qubit Wheel of Fig. 4. Dropping one of the qubits from all the observables would not leave a valid proof because two of the spokes of the wheel would then no longer be IDs.

\section{\label{sec:4}Parity proofs based on the three-qubit Pauli group}

We now show how the diagrams of the previous section can be used to obtain parity proofs of the KS theorem. We do this in a few cases only, but the general procedure should then be clear. We begin by reviewing the proof of Kernaghan and Peres \cite{KP1995} and showing that the 40-ray set they considered has two new parity proofs in it besides the one they found. Then we discuss the parity proofs yielded by two of our other diagrams, both of which have their points of interest.

{\bf \subsection{\label{subsec:Pentagram}The GHZ-Mermin Pentagram}}

Kernaghan and Peres used the GHZ-Mermin pentagram (left diagram of Fig.3) to obtain a set of 40 rays from which they extracted a parity proof of the KS theorem. Table 1 explains how they obtained their rays. Each row of the table shows the observables in one of the ID4s of the pentagram, followed by their eight simultaneous eigenstates, with the eigenstates labeled (at the tops of the columns) by their eigenvalue signatures with respect to these observables. The eigenstates (or rays) have been numbered 1 to 40, with the numbering chosen to be identical to that of Kernaghan and Peres \cite{KP1995}.\\

The information in Table 1 allows all the orthogonalities among the rays to be worked out according to the following rule: each ray is orthogonal to those (and only those) that have the opposite eigenvalue signature(s) from it for the observable(s) they have in common in their defining ID4s (if the ID4s have no observables in common, then the rays are not orthogonal). The basis table of the rays is then easily obtained, and is shown in Table 2 (by a ``basis" we mean a set of mutually orthogonal rays, or projectors, that spans the space). The basis table has the property of being ``saturated" (i.e., all orthogonalities are represented in it) and it is also highly symmetrical in that each ray is orthogonal to 23 others and occurs in exactly five bases. The bases themselves are of two types that we will term ``pure" and ``hybrid". The pure bases are just the eigenbases of Table 1, while the hybrid bases each consist of an equal mixture of rays from a pair of pure bases. The relationship between the pure and hybrid bases can be grasped by looking at the ``hybridization" diagram of Fig. 9: the five pure bases are represented by the dots at the vertices of the pentagon, and the twenty hybrids by the dots at the ends of the lines passing through every pair of vertices (or pure bases). For example, if the pure bases 1 2 3 4 5 6 7 8 and 9 10 11 12 13 14 15 16 are associated with two of the vertices, then the hybrids 1 2 3 4 13 14 15 16 and 5 6 7 8 9 10 11 12 at the ends of the line segment passing through them each get half of their rays from each of these pure bases.\\

\begin{table}[ht]
\centering 
\begin{tabular}{|c | c c c c c c c c |} 
\hline 
 Observables & ++++  &   ++ -- -- &  + -- + --  &  + -- -- +& -- ++ -- &   -- + -- + &  -- -- + + &  -- -- -- --  \\
\hline
$ZII$, $IZI$, $IIZ$ , $ZZZ$           & $1$ & $2$ & $3$ & $4$ & $5$ & $6$ & $7$ & $8$ \\
$ZII$, $IXI$, $IIX$ , $ZXX$           & $9$ & $11$ & $10$ & $12$ & $13$ & $15$ & $14$ & $16$ \\
$XII$, $IZI$, $IIX$ , $XZX$           & $17$ & $19$ & $21$ & $23$ & $18$ & $20$ & $22$ & $24$ \\
$XII$, $IXI$, $IIZ$ , $XXZ$           & $25$ & $29$ & $27$ & $31$ & $26$ & $30$ & $28$ & $32$ \\
$ZZZ$, $ZXX$, $XZX$ , $-XXZ$          & $33$ & $35$ & $34$ & $36$ & $40$ & $37$ & $38$ & $39$ \\
\hline
\end{tabular}
\caption{The 40 rays of Kernaghan and Peres, derived from the GHZ-Mermin pentagram (left diagram of Fig. 3). The rays are identified by their eigenvalue signatures with respect to the observables at their left and numbered 1 to 40 following the scheme of Kernaghan and Peres.}
\label{table1} 
\end{table}

\begin{table}[ht]
\centering 
\begin{tabular}{|c | c c c c c c c c |} 
\hline 
Index &\multicolumn{8}{|c|}{Rays in Basis}\\
\hline
1 & 1 & 2 & 3 & 4 & 5 & 6 & 7 & 8 \\
2 & 9 & 10 & 11 & 12 & 13 & 14 & 15 & 16 \\
3 & 17 & 18 & 19 & 20 & 21 & 22 & 23 & 24 \\
4 & 25 & 26 & 27 & 28 & 29 & 30 & 31 & 32 \\
5 & 33&34&35&36&37&38&39&40 \\
\hline
6&1&2&3&4&13&14&15&16\\
7&1&2&5&6&21&22&23&24\\
8&1&3&5&7&29&30&31&32\\
9&1&4&6&7&37&38&39&40\\
10&2&3&5&8&33&34&35&36\\
11&2&4&6&8&25&26&27&28\\
12&3&4&7&8&17&18&19&20\\
13&5&6&7&8&9&10&11&12\\
14&9&10&13&14&19&20&23&24\\
15&9&11&13&15&27&28&31&32\\
16&9&12	&14	&15	&34	&36	&38	&39\\
17&10&	11&	13&	16&	33&	35&	37&	40\\
18&10&	12&	14&	16&	25&	26&	29&	30\\
19&11&	12&	15&	16&	17&	18&	21&	22\\
20&17&	19&	21&	23&	26&	28&	30&	32\\
21&17&	20&	22&	23&	35&	36&	37&	39\\
22&18&	19&	21&	24&	33&	34&	38&	40\\
23&18&	20&	22&	24&	25&	27&	29&	31\\
24&25&	28&	30&	31&	33&	36&	37&	38\\
25&26&	27&	29&	32&	34&	35&	39&	40\\
\hline
\end{tabular}
\caption{The 25 bases formed by the 40 rays of Kernaghan and Peres. The five pure bases are shown first, followed by the 20 hybrid bases. The bases have been numbered from 1 to 25 for later convenience.}
\label{table2} 
\end{table}

\begin{figure}[htp]
\begin{center}
\includegraphics[width=.40\textwidth]{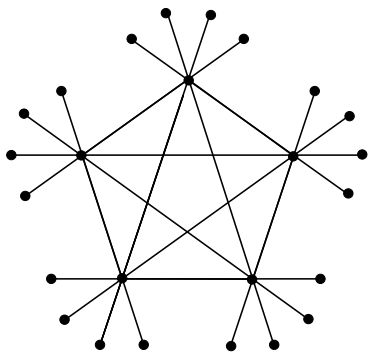}
\end{center}
\caption{Hybridization pattern of the bases in the 40-ray Kernaghan-Peres set. The dots at the vertices of the pentagon represent the five ``pure'' bases and the dots at the ends of the lines passing through pairs of pure bases the hybrids they give rise to.}
\label{Fig.9}
\end{figure}

A parity proof of the KS theorem can be obtained by picking out suitable subsets of bases from Table 2. But before doing that we discuss the notion of a parity proof in general. A set of $R$ rays and $B$ bases will be said to provide a parity proof of the KS theorem if (a) $B$ is odd, and (b) each of the $R$ rays occurs an even number of times among the $B$ bases. These two conditions guarantee a KS proof because it is impossible for a noncontextual hidden-variables theory to assign a 0 or a 1 to each of the rays in such a way that each basis contains exactly one ray assigned the value 1. This conclusion follows from a simple parity argument, and so we refer to this type of proof as a ``parity proof". Instead of talking of the rays in a parity proof, we can talk of the equivalent projectors. There is also no need for the projectors to be of rank-1, but they can be of higher rank and moreover need not all be of the same rank. All that is needed for a KS proof is that one have a set of projectors that form an odd number of bases, with each projector occurring in an even number of these bases. We will actually encounter this more general type of situation in the two cases to be discussed later.\\

Kernaghan and Peres showed how to pick out a set of 11 bases from Table 2 that gives a parity proof of the KS theorem. We have found that it is also possible to pick out sets of 13 and 15 bases that give parity proofs. One example of each of these three types of proofs is given in Table 3, which also summarizes their principle characteristics. The first column of Table 3 lists the number of rays (R) and bases (B) in each of the proofs and the second column goes on to specify the multiplicities of all the rays occurring in a proof. For example, the symbol $28_{2}8_{4}$-$11_{8}$ for the first proof indicates that it has 28 rays that each occur twice and 8 rays that each occur four times among its 11 bases (with the subscript 8 simply indicating the number of rays in a basis). The third column lists the bases in a proof of each type, using the numbering scheme of Table 2, and the fourth column lists the number of proofs of each type. It is interesting that the total number of parity proofs of all three types is $1024 = 2^{10}$, where 10 is the number of hybrid bases in each of the proofs. This is similar to what we have observed in two other cases: the 24-24 set of rays and bases of Peres has $2^{9} = 512$ parity proofs \cite{Waegell2011b} and a 40-40 set we discovered recently\cite{Waegell2011} has $2^{15} = 32768$ parity proofs, where again the exponents 9 and 15 are the number of hybrid bases in these proofs. 

\begin{table}[ht]
\centering 
\begin{tabular}{|c |c |c |c |} 
\hline 
 R-B & Symbol  &   Example Proof  &  Number  \\
\hline
36-11 & $28_{2}8_{4}$-$11_{8}$ & 1 6 7 8 10 14 15 17 20 21 25& 320 \\
38-13 & $24_{2}14_{4}$-$13_{8}$ & 1 2 3 6 7 8 10 14 15 16 20 22 25& 640 \\
40-15 & $20_{2}20_{4}$-$15_{8}$ & 1 2 3 4 5 6 7 8 10 14 15 16 20 22 24& 64 \\
\hline
\end{tabular}
\caption{Parity proofs in the 40-25 set of Kernaghan and Peres. See text for explanation.}
\label{table3} 
\end{table}

{\bf \subsection{\label{subsec:3Square} The three-qubit Peres-Mermin square}}

We next look at the three-qubit Peres-Mermin square of Fig. 4, which is a generalization of the two-qubit square of Fig.1. The simultaneous eigenstates of each of the ID3s in this diagram define four mutually orthogonal rank-2 projectors, and all six ID3s give rise to 24 projectors that are identified by their eigenvalue signatures, and numbered from 1 to 24, in Table 4. The explicit forms of these projectors are not needed, but we nevertheless give one example of them for the sake of the interested reader. Table 5 shows the two-dimensional subspaces corresponding to the projectors defined by the observables in the fourth row of Table 4. We next want to determine all the orthogonalities between the projectors and all the bases formed by them. As explained in Sec. \ref{subsec:Pentagram}, this can be done merely by comparing the eigenvalue signatures of different projectors. Two projectors are orthogonal if and only if their eigenvalue signatures differ for at least one observable of which they are both eigenstates.\\

In this way it is easy to see that the 24 projectors form 24 bases, of four projectors each, with each projector occurring in four bases. This is identical to the situation that occurs in the two-qubit Peres-Mermin square, the only difference being that rank-2 (rather than rank-1) projectors are involved in the present case. However all the details of the parity proofs are identical in the two cases. The numbering of the projectors in Table 4 has been chosen to make all the parity proofs in the present case identical to those for the two-qubit Peres-Mermin square listed in \cite{Waegell2011b}. For convenience, we summarize the details of all the parity proofs in this system in Table 6, using the same format as in Table 3. It is interesting that the most compact parity proof contained in this system, which involves 18 projectors and 9 bases, is more economical than the proof of Kernaghan and Peres (which involves 36 projectors and 11 bases). It might be thought that this proof is a trivial extension of the 18-9 proof of Cabello et al \cite{Cabello1996}, but it is not, because it cannot be reduced to that proof by ignoring all the measurements on any one of the qubits.

\begin{table}[ht]
\centering 
\begin{tabular}{|c | c c c c |} 
\hline 
 Observables & +++  &   + -- -- &  -- + -- &  -- -- + \\
\hline
$ZIZ$, $ZZI$, $IZZ$            & $1$ & $2$ & $3$ & $4$ \\
$XIX$, $XXI$, $IXX$            & $5$ & $6$ & $7$ & $8$ \\
$YIY$, $YYI$, $IYY$            & $9$ & $10$ & $11$ & $12$ \\
$ZIZ$, $XIX$, $-YIY$            & $13$ & $14$ & $15$ & $16$  \\
$ZZI$, $XXI$, $-YYI$            & $19$ & $20$ & $17$ & $18$  \\
$IZZ$, $IXX$, $-IYY$            & $22$ & $21$ & $24$ & $23$  \\
\hline
\end{tabular}
\caption{The 24 rank-2 projectors defined by the six ID3s of the three-qubit Peres-Mermin square of Fig.4. The projectors in each row are the simultaneous eigenstates of the observables to their left, with the eigenvalue signatures indicated above them.}
\label{table4} 
\end{table}

\begin{table}[ht]
\centering 
\begin{tabular}{|c | c |} 
\hline 
 Projector & 2-d subspaces \\
\hline
$13$            & $(1,0,0,0,0,1,0,0), (0,0,1,0,0,0,0,1)$  \\
$14$            & $(1,0,0,0,0,-1,0,0), (0,0,1,0,0,0,0,-1)$  \\
$15$            & $(0,1,0,0,1,0,0,0), (0,0,0,1,0,0,1,0)$  \\
$16$            & $(0,1,0,0,-1,0,0,0), (0,0,0,1,0,0,-1,0)$  \\
\hline
\end{tabular}
\caption{Pairs of linearly independent vectors defining the two-dimensional subspaces corresponding to projectors 13-16 of Table 4. (The vector $(a,b,\cdots,h)$ represents the state $a|000\rangle + b|001\rangle + \cdots + h|111\rangle$, where $|000\rangle, |001\rangle$ etc. are the computational basis states.)}
\label{table5} 
\end{table}

\begin{table}[ht]
\centering 
\begin{tabular}{|c |c |c |} 
\hline 
 R-B & Symbol  &  Number  \\
\hline
18-9 & $18_{2}$-$9_{4}$ &  16 \\
20-11 & $18_{2}2_{4}$-$11_{4}$ & 240 \\
22-13 & $18_{2}4_{4}$-$13_{4}$ & 240 \\
24-15 & $18_{2}6_{4}$-$15_{4}$ & 16 \\
\hline
\end{tabular}
\caption{Parity proofs in the three-qubit Peres-Mermin square.}
\label{table5} 
\end{table}

{\bf \subsection{\label{subsec:3Kite} The three-qubit Kite}}

The parity proofs of the last two subsections have involved projectors of only a single rank (either rank-1 or rank-2) and come in a relatively small number of varieties (three in one case and four in the other). However the proofs yielded by many of our diagrams are considerably more complex. To give some idea of this, we now discuss the proofs contained in the kite diagram of Fig. 6, which consists of four ID3s and two ID4s. Each ID3 gives rise to four rank-2 projectors and each ID4 to eight rank-1 projectors, and the whole diagram to 32 projectors that form 36 bases in all. We can characterize this system of projectors and bases by the brief symbol 32-36 or the detailed symbol $16_{10}{\bf 16_{4}}$-$16_{8}8_{6}12_{4}$, which indicates that there are 16 projectors of multiplicity 10 and 16 of multiplicity 4, and that they form 16 bases of size eight, 8 bases of size six and 12 bases of size four. We have indicated the rank-2 projectors in boldface and the rank-1 projectors in ordinary type. The varying basis sizes arise because the bases can now consist of different mixes of rank-1 and rank-2 projectors; the bases of size 8 consist entirely of rank-1 projectors, those of size 4 entirely of rank-2 projectors and those of size 6 of four rank-1 and two rank-2 projectors. This 32-36 system has a large number of parity proofs in it, some of which are listed in Table 7. The first column lists the number of projectors (P) and bases (B) in each proof, the second the detailed symbol of the proof (using the same conventions as explained in connection with the 32-36 system) and the third the number of distinct copies (or ``replicas") of this proof in the entire system. It is worth noting that the briefest proof in this system, which involves 24 projectors and 9 bases, is more economical than the proof of Kernaghan-Peres.

\begin{table}[ht]
\centering 
\begin{tabular}{|c |c |c |} 
\hline 
 P-B & Detailed Symbol & Count \\
\hline
24-9 & $12_{2}{\bf 12_{2}}$-$1_{8}4_{6}4_{4}$ & 16 \\
26-11 & $8_{2}6_{4}{\bf 12_{2}}$-$3_{8}4_{6}4_{4}$ & 96\\
28-13 & $4_{2}12_{4}{\bf 12_{2}}$-$5_{8}4_{6}4_{4}$ & 192 \\
30-15 & $4_{2}12_{4}{\bf 12_{2}2_{4}}$-$5_{8}4_{6}6_{4}$ & 1152\\
32-17 & $4_{2}12_{4}{\bf 12_{2}4_{4}}$-$5_{8}4_{6}8_{4}$ & 192 \\
\hline
\end{tabular}
\caption{Some parity proofs in the three-qubit kite diagram.}
\label{table6} 
\end{table}

A quick check on the consistency of the detailed symbols in Table 7 is to note that the total number of projectors and bases in them match the numbers in the first column. A slightly more involved check is to count the total number of projectors of each rank in both halves of the symbol and to verify that the two counts agree. Reassurance that the symbols represent parity proofs comes from the fact that all the subscripts in the first half of the symbol are even (implying that each projector occurs an even number of times among the bases) and the total number of bases in the second half is odd. We do not exhibit any of the proofs here, in order to save space, but hope their detailed symbols will give the reader some feeling for what they are like. We should add that the listing in Table 7 is only partial, and that there are actually 33 different types of proofs in this system. Each proof has many replicas under symmetry, and the total number of proofs in this system is 33152. The parity proofs yielded by our other diagrams have the same general features.

\section{\label{sec:5} Concluding remarks}

We have presented a family of new proofs of the KS theorem based on a system of three qubits. Our proofs are of two types: observable-based proofs, based on commuting sets of observables from the three-qubit Pauli group, and parity proofs, based on sets of projectors and bases derived from the former proofs. Although our proofs have been presented as state-independent proofs of the KS theorem, it is worth pointing out that they can be converted into proofs of Bell's theorem ``without probabilities" if some additional resources are granted. The way this can be done is the following \cite{note}: (1) a source repeatedly emits three singlet states, with one member of each singlet going to Alice and the other to Bob in every run; (2) in any run Alice and Bob each measure a randomly chosen set of commuting observables from one of our KS proofs on their three qubits and verify, over a series of many runs, that the KS theorem is valid; (3) finally Alice and Bob get together and find that in any run in which they measured one or more of the same observables, those observables always had the same values. Since Alice and Bob are spacelike separated, the last observation helps replace the questionable assumption of noncontextuality in the proof of the KS theorem by the more respectable assumption of locality and thus  promotes it into a proof of Bell's theorem ``without probabilities". Of course this is only a gedanken demonstration because it assumes that the singlets are uncorrupted by noise and that the detectors are perfect, but it is interesting that it can be given all the same.\\

It is our hope that the new proofs of the KS theorem we have presented here can contribute to a greater understanding of quantum contextuality and also find applications in quantum information protocols.


\end{document}